\documentclass[a4paper,12pt]{article}
\usepackage[english]{babel}
\usepackage{xcolor}
\usepackage{amsmath}
\usepackage{amsfonts}
\usepackage{amssymb}
\usepackage{calrsfs}
\usepackage{graphicx}
\usepackage{bussproofs}
\usepackage{proof}
\usepackage{latexsym}
\usepackage{titling}
\usepackage[ND,SEQ]{prftree}
\usepackage{multicol, caption}
\usepackage{setspace}
\usepackage[nottoc,numbib]{tocbibind}
\usepackage{pdflscape}
\usepackage{stmaryrd}
\usepackage{wasysym}
\usepackage[natbibapa]{apacite}
\usepackage[left=30mm, right=30mm, top=20mm, bottom=20mm]{geometry}

\begin{document}

\title{What are acceptable reductions? \\{\large Perspectives from proof-theoretic semantics and type theory}}


\author{Sara Ayhan\thanks{I would like to thank David Ripley, Ansten Klev, Will Stafford and Luca Tranchini for reading early versions of this paper and giving me extensive feedback on it, which immensely helped me to clarify my ideas. Thanks also go to an anonymous referee for their very constructive and helpful report. I am especially grateful
to Heinrich Wansing for his thorough reading and feedback on the paper and his kind and valuable suggestions for improvement.}
\\
{\small Ruhr University Bochum, Institute of Philosophy I}\\
{\small sara.ayhan@rub.de}}

\date{}

    \maketitle
    \begin{abstract}
        It has been argued that reduction procedures are closely connected to the question about identity of proofs and that accepting certain reductions would lead to a trivialization of identity of proofs in the sense that every derivation of the same conclusion would have to be identified. In this paper it will be shown that the question, which reductions we accept in our system, is not only important if we see them as generating a theory of proof identity but is also decisive for the more general question whether a proof has meaningful content. There are certain reductions which would not only force us to identify proofs of \emph{different arbitrary formulas} but which would render derivations in a system allowing them \emph{meaningless}. To exclude such cases, a minimal criterion is proposed which reductions have to fulfill to be acceptable.
    \end{abstract}

%%================================%%
%% Sample for structured abstract %%
%%================================%%

\section{Introduction} What are acceptable reductions\footnote{Note that I am focussing strictly on \emph{reductions} in this paper, i.e., procedures that cut out what is in some way considered a detour of a derivation, not conversions in general, like expansions or permutations. The latter are certainly also of great interest for proof-theoretic semantics but would extend the scope of this paper.} in the context of proofs and why is it important to distinguish these from `bad' ones? 
As Schroeder-Heister and Tranchini \citeyearpar[p. 574]{PSHT2017} argue, from a philosophical point of view, or more specifically a standpoint of proof-theoretic semantics \cite[see][]{sep-proof-theoretic-semantics}, reduction procedures are closely connected to the question about identity of proofs:
If we take proofs to be abstract entities represented by (natural deduction) derivations, then derivations belonging to the same equivalence class induced by the reflexive, symmetric, and transitive closure of reducibility can be said to represent the same proof object.\footnote{This goes back to Prawitz \citeyearpar[pp. 257-261]{Prawitz1971}, who credits the idea to Per Martin-L\"of; many others have defended this view since.}
As they show, accepting certain reductions, more specifically accepting the so-called \emph{Ekman-reduction} (see below), would lead to a trivialization of identity of proofs in the sense that every derivation of the same conclusion would have to be identified.
They suggest such a trivialization as a criterion to disallow reductions.   
I will argue that the question, which reductions we accept in our system, is not only important if we see them as generating a theory of proof identity but is also decisive for the more general question whether a proof has meaningful content, i.e., it does not only matter to the question about the \emph{denotation} of proofs but also to the question about their \emph{sense}.
Therefore, we need to be careful: We cannot just accept any reduction, i.e., any procedure eliminating some kind of detour in a derivation.

An example of a reduction not belonging to the usual reductions is Ekman-reduction, as it is presented in \citep{Ekman1994, Ekman1998} and extensively discussed by Schroeder-Heister and Tranchini \citeyearpar{PSHT2017, PSHT2018} and Tennant \citeyearpar{Tennant2021}:
\vspace{0.2cm}

\quad  
\infer[\scriptstyle\rightarrow E]{A} 
{\;\;\;\;\;\;\;\;\;\;\;\;{B \rightarrow A} \quad \infer[\scriptstyle\rightarrow E]{B}{\;\;{A \rightarrow B} \quad \quad {\infer*{A}{\mathcal{D}}}}}
\quad
$\rightsquigarrow$
\quad
\infer*{A}{\mathcal{D}}
\vspace{0.3cm}

What we want in light of such a non-standard kind of reduction are criteria determining which reductions can be allowed in our system and which should be dismissed.
It is advantageous for such an approach to exploit the so-called \emph{Curry-Howard correspondence} \cite[see, e.g.,][]{SU} and examine proof systems annotated with $\lambda$-terms.
These make the structure of our derivations explicit and facilitate to show what is wrong with potential reductions and why they should not be admitted in our system.
The question then shifts to asking which reduction procedures for \textit{terms} can be allowed.
The $\lambda$-calculus and some well-known properties thereof can provide us with directions as to what could be (un)desirable features of reductions.

Annotating Ekman-reduction with terms then shows something else - besides Schroeder-Heister and Tranchini's point - that is essentially problematic about this reduction.
Indeed, I want to show that allowing it would be equal to allowing a reduction for \texttt{tonk}, i.e., a reduction for a derivation consisting of a \texttt{tonk}-introduction rule followed by its elimination rule (see below, section 3.1.1).
It is generally agreed upon, though, that there cannot be a sensible reduction for this connective. 
By creating a \texttt{tonk}-reduction in the same fashion as it would be done for other connectives, the consequences of allowing this reduction are made explicit and it can be shown that those would be the same for Ekman-reduction: Not only would these reductions induce an equivalence relation relating different terms in normal form of the \emph{same} conclusion, they would also allow to reduce a term of one type to a term of an \emph{arbitrary} other.
If we take reductions as generating identity between proofs, then that would also force us to identify proofs of arbitrarily different formulas.
But even if we reject this assumption (some researchers do not find this theory of proof identity very compelling, see section 4), I will argue that allowing such reductions would render derivations in such a system \emph{meaningless}.

\section{Reduction procedures in natural deduction and $\lambda$-calculus}
The reductions for the connectives of, say, minimal propositional logic, corresponding to $\beta$-reductions in $\lambda$-calculus, are meant to eliminate unnecessary detours of the following form: There is a formula, called \emph{maximal formula}, which is both the conclusion of an application of an introduction rule of a connective as well as the major premise of an applied elimination rule governing the same connective.
It can be shown for those connectives (see below for $\rightarrow$) that in these cases the maximal formula (below $A \rightarrow B$) can be eliminated without losing anything essential because, as Prawitz \citeyearpar[p. 251]{Prawitz1971} argues, this procedure is just a way to make the \emph{inversion principle}\footnote{The principle, to which (at least) the rules governing the connectives of minimal logic adhere, saying that nothing new is obtained by an elimination immediately following an introduction of the major premise of the elimination rule \citep{Prawitz1971}.} explicit.

\begin{center}

\quad  
\infer[\scriptstyle\rightarrow E]{B} 
{{\infer[\scriptstyle\rightarrow I]{A \rightarrow B}{\infer*[\scriptstyle\mathcal{D}]{B}{[A]}}} \quad \infer*{A}{\mathcal{D'}} }
\quad
$\rightsquigarrow$
\quad
\infer*[\scriptstyle\mathcal{D}]{B}{\infer*{[A]}{\mathcal{D'}}}
\end{center}

\noindent It can and has been argued, however, that there are more reductions than the ones for the connectives that are usually considered \cite[see, e.g.,][]{Tennant1995}.
One of those, presented in \citep{Ekman1994, Ekman1998}, will be discussed in this paper.

As mentioned above, in using a term-annotated proof system we are implementing the Curry-Howard correspondence, which takes the view of proofs as programs and formulas as types as a basis and which states a close correspondence between these notions in the simply typed $\lambda$-calculus and natural deduction (ND) systems of intuitionistic logic.
For our purposes here it suffices to consider the $\rightarrow$-fragment of intuitionistic logic and correspondingly the system $\lambda ^{\rightarrow}$.
We use $\rho$, $\sigma$, $\tau$,... for arbitrary atomic formulas, $A$, $B$, $C$,... for arbitrary formulas, and $\Gamma$, $\Delta$,... for sets of formulas.
The concatenation $\Gamma$, $A$ stands for $\Gamma \cup \left\{A\right\}$.
For term variables, $x$, $y$, $z$,... are used and $r$, $s$, $t$,... for arbitrary terms.
Furthermore, we use `$\equiv$' to denote \textit{syntactic identity} between terms, types, or derivations.
The following are our term-annotated ND-rules with the corresponding $\beta$-reduction:

\vspace{0.3cm}

\quad  
\infer[\scriptstyle\rightarrow I]{\lambda x.t: A \rightarrow B} 
{\infer*{t : B}{{[x: A]} }}
\quad  
\infer[\scriptstyle\rightarrow E]{App(s, t): B} 
{\infer*{s: A \rightarrow B}{{\Gamma}} \quad \infer*{t : A}{{\Delta} }}
\quad
$App(\lambda x.t, s) \rightsquigarrow_{\beta} t[s/x]$
\vspace{0.3cm}

We read $t : A$ as ``term $t$ is of type $A$'' or, in the `proof-reading', ``$t$ is a proof of formula $A$".
Such an expression is also called a \textit{type assignment statement} with the $\lambda$-term being the \textit{subject} and the type the \textit{predicate}.
Thus, we use a type system \textit{\`{a} la Curry} here, in which the terms are not typed, in the sense that the types are part of the term's structure, but are \textit{assigned} types according to type assignment rules, which in our case are simply the rules above.
With $t : A$ we express that $A$ is the \emph{principal type} of term $t$, i.e., the most general type that can be assigned to $t$.\footnote{For example, for the term $\lambda x.x$, its types could be $p \rightarrow p$, $q \rightarrow q$, $(p \rightarrow q) \rightarrow (p \rightarrow q) $, etc., while its principal type would be $A \rightarrow A$.}
Substitution is expressed by $t[s/x]$, meaning that in term $t$ every free occurrence of $x$ is substituted with $s$.
The usual capture-avoiding requirements for variable substitution are to be observed.
I will follow standard terminology of type theory here and call a term of the form $App(\lambda x. t,s)$ a $\beta$\textit{-redex} and the corresponding term $t[s/x]$ its \textit{contractum}.\footnote{For the following definitions (with only slightly differing formulations and notations), see \citep{Barendregt1992}, \citep{Girard1989}, and \citep{HS2008}. I will use the same terminology (without the `$\beta$-') for any reduction procedures, whether or not they will be found acceptable in the course of the paper. For reduction procedures in general, see also \citep{BaaderNipkow}.}
Replacing an occurrence of a $\beta$-redex contained in a term by its contractum is called a $\beta$-\textit{contraction} and if there is a finite (possibly empty) series of $\beta$-contractions changing term $t$ to $t'$, we say that $t$ $\beta$-\textit{reduces} to $t'$ and write $t$ $\rightsquigarrow _{\beta}$ $t'$.
The reduction relation $\rightsquigarrow _{\beta}$ is reflexive and transitive and closed under $\alpha$-conversion, i.e., renaming of bound variables.
A term that contains no $\beta$-redexes is said to be $\beta$\textit{-normal} or in $\beta$\textit{-normal form ($\beta$-nf)} and $t'$ is the $\beta$-nf of $t$ if $t \rightsquigarrow _{\beta} t'$ and $t'$ is $\beta$-normal.

In general (if we consider more connectives than $\rightarrow$), the correspondence to the introduction and elimination rules in $\lambda$-calculus is that each connective has its own \emph{constructor}, an operator constructing canonical objects of particular types, and a \emph{destructor} specifying the use of these objects in computations.
A $\beta$-redex consists of a destructor applied to the constructor of the same connective, so Curry-Howard correspondence always gives us an analogy between a $\beta$-redex and a proof detour consisting of an elimination immediately following the introduction of the same connective \citep[p. 87f.]{SU}.

It is important to stress that reduction procedures can be interpreted in two different (yet certainly closely related) ways (see, e.g., \cite[Ch. 2.3]{Barendregt1992}, \cite[pp. 18-20]{Girard1989}, \cite[pp. 11-18]{HS2008}).
One interpretation is to see reductions as inducing an identity relation, i.e., on this view the relation applies equally in both directions.
We will speak of $\beta$-\textit{equality} of terms in these cases and use $=_{\beta}$ to express this relation.
It is just like, e.g., $App(\lambda x.x ^{2}+7,2)$ $\rightsquigarrow _{\beta}$ 11 expresses the fact that 4+7=11.

Another interpretation of reductions is to see them as \textit{directed} computations, calculations, or executions corresponding to the idea of program evaluation.
On this view, the asymmetry between redex and contractum must be stressed:
In our example `4+7' can be interpreted as `11' by doing a calculation. 
A reduction procedure is seen as an evaluation that is run on a term and thereby interprets this term in a different way.
The non-symmetric $\beta$-reducibility relation implies the symmetric relation of $\beta$-equality but not the other way around \cite[p. 16]{HS2008}.
Hence, if $t$ $\rightsquigarrow _{\beta}$ $t'$, then $t$ $= _{\beta}$ $t'$; but not: if $t$ $= _{\beta}$ $t'$, then $t$ $\rightsquigarrow _{\beta}$ $t'$.
Just like 4+7 evaluates to 11 but not the other way around: 11 is fully evaluated; it is already in normal form, i.e., we do not \textit{reduce} it to 4+7.

One of the most important results in $\lambda$-calculus, which will also be important for this paper, is the so-called \textit{Church-Rosser Theorem} stating the \textit{confluence property} for $\beta$-reduction: 
\begin{description}
\item[\textbf{Church-Rosser Theorem:}]  If a term can be reduced to two syntactically different terms, then there is a term to which these two can be reduced.
Put formally, if $t$ $\rightsquigarrow _{\beta}$ $t'$ and $t$ $\rightsquigarrow _{\beta}$ $t''$, then there is a term $s$ such that $t'$ $\rightsquigarrow _{\beta}$ $s$ and $t''$ $\rightsquigarrow _{\beta}$ $s$.
\end{description}

Likewise, this property holds for $\beta$-equality: if two syntactically different terms are $\beta$-equal, then there is a term to which they both can be reduced in finitely many steps, i.e., if $t$ $= _{\beta}$ $t'$ and $t \not\equiv t'$, then there is a term $s$ such that $t$ $\rightsquigarrow _{\beta}$ $s$ and $t'$ $\rightsquigarrow _{\beta}$ $s$.
One corollary of this is the uniqueness of $\beta$-normal forms for terms (provided they have a normal form).
Another important corollary for our purposes is that two terms $t$ and $t'$ that are in $\beta$-nf \textit{and} syntactically distinct cannot be $\beta$-equal, which means that the relation $= _{\beta}$ is non-trivial: not all terms are $\beta$-equal \cite[p. 17]{HS2008}.\footnote{This also implies the consistency of the simply typed $\lambda$-calculus, see \cite[Ch. 2.3]{Barendregt1992} and \cite[p. 23]{Girard1989}.}

\section{What distinguishes `good' from `bad' reductions?}
\subsection{Problematic reductions}
\subsubsection{\texttt{Tonk}-reduction}
The reduction procedure considered above for $\rightarrow$ for eliminating maximal formulas, that arise from applying an elimination rule immediately after the corresponding introduction rule, works equally well for our other `well-behaved' connectives \citep{Prawitz1965}.
A comparison with the notorious connective \texttt{tonk} might help to see, however, why this is not the case for every connective. 
\texttt{Tonk} was introduced by Prior \citeyearpar{Prior} as an ad absurdum-attack on the idea of proof-theoretic semantics\footnote{Although this specific term has been introduced only later in 1991 by Schroeder-Heister \citeyearpar{sep-proof-theoretic-semantics}, the general idea has been prevalent much longer.}: If it was only the rules giving the meaning of a connective, no other metaphysically underlying concept, then what would stop the proof-theoretic semanticist from accepting the following rules?

\begin{center}

\quad  
\infer[\scriptstyle \texttt{tonk} I]{A~ \texttt{tonk}~ B}{A} 
\quad \quad \quad \quad  
\infer[\scriptstyle \texttt{tonk} E]{B}{A ~\texttt{tonk}~ B} 
\end{center}

Applying these immediately after each other gives us a derivation from arbitrary $A$ to arbitrary $B$, i.e., our system would trivialize.
Additionally, there is no real way to make out a reasonable reduction procedure in this case, which is, of course, due to the fact that \texttt{tonk} violates the inversion principle.
This has been one of the ways to give a reason why \texttt{tonk} can be considered inadmissible.\footnote{Since the \texttt{tonk}-rules do not adhere to the inversion principle, they are not in \textit{harmony}. On this notion as a criterion for acceptable connectives, see, e.g., \citep{Tennant1978}, \citep{Dummett1991}, \citep{Read2010}, \citep{FrancezDyckhoff}, and \citep{Tranchini2015}.}
It should be noted that there are approaches to \texttt{tonk}, which do not even consider it inadmissible in principle but which rather question our underlying assumptions about logical consequence on the grounds of which we dismiss \texttt{tonk}.\footnote{Cook \citeyearpar{Cook2005} and Ripley \citeyearpar{Ripley2015}, e.g., argue like this in claiming that if we do not assume a transitive consequence relation, then an extension with \texttt{tonk} would not yield inconsistency. See also \citep{Wansing2006}, however, where it is shown that the problems of \texttt{tonk} avoided in a non-transitive system can be recreated by other \texttt{tonk}-like connectives.}
Yet, I want to emphasize that even with an argumentation that accepts \texttt{tonk}, to my knowledge, there is still no way of giving an acceptable reduction procedure for this connective.

Can this be made explicit with term annotations?
Leaving $\lambda$-calculus, we can still give term-annotated rules and a corresponding reduction for non-standard connectives, like for a Liar-connective $L$ for example, as it has been proposed in \citep{PSH2012c}:

\begin{center}

\quad  
\infer[\scriptstyle \texttt{L} I]{lt: L}{t:L \rightarrow \bot} 
\quad \quad \quad \quad 
\infer[\scriptstyle \texttt{L} E]{l't: L \rightarrow \bot}{t:L} 
\quad \quad \quad \quad 
$l'lt \rightsquigarrow_{\texttt{L}} t$
\end{center}

We have $l$ here serving as a constructor for the introduction rule and $l'$ as a destructor in the elimination rule.
We can do the same for \texttt{tonk}, annotating the rules with a constructor $k$ and a destructor $k'$:

\begin{center}

\quad  
\infer[\scriptstyle \texttt{tonk} I]{kt: A~ \texttt{tonk}~ B}{t:A} 
\quad \quad \quad \quad 
\infer[\scriptstyle \texttt{tonk} E]{k't: B}{t:A ~\texttt{tonk}~ B} 
\end{center}

Just like for $L$ a non-normal term for \texttt{tonk} would then be constructed by applying the destructor to the constructor, which is, as for the usual connectives, the result of a derivation containing the conclusion of the introduction rule as the major premise of the elimination rule, i.e.:

\begin{center}

\quad  
\infer[\scriptstyle \texttt{tonk} E]{k'kt: B}{\infer[\scriptstyle \texttt{tonk} I]{kt:A ~\texttt{tonk}~ B}{t:A}}
\end{center}

The usual reduction would be to reduce the term for the conclusion of the elimination rule to the one of the premise of the introduction rule, so analogous to the Liar-reduction: $k'kt \rightsquigarrow_{\texttt{tonk}} t$.
However, $t$ is assigned type $A$, while $k'kt$ is assigned $B$.
So, if we would accept this reduction, it would mean to accept a reduction relating terms of arbitrarily different types.
In the following I want to show that what is wrong with Ekman-reduction is essentially the same as in the case of \texttt{tonk}-reduction and on this basis identify  what could be a good criterion for reductions of proofs in terms of type theory.

\subsubsection{Ekman-reduction}
Again, Ekman-reduction has the following form:

\quad  
\infer[\scriptstyle\rightarrow E]{A} 
{\;\;\;\;\;\;\;\;\;\;\;\;{B \rightarrow A} \quad \infer[\scriptstyle\rightarrow E]{B}{\;\;{A \rightarrow B} \quad \quad {\infer*{A}{\mathcal{D}}}}}
\quad
$\rightsquigarrow$
\quad
\infer*{A}{\mathcal{D}} 

\vspace{0.3cm}

The motivation for Ekman to consider this reduction was to give a counterexample to Tennant's \citeyearpar{Tennant1982} proof-theoretic characterization of paradoxes.
According to this, a paradoxical derivation is one that yields a non-normalizable derivation of $\bot$.
Tennant considers several examples, like versions of the Liar paradox, Curry's paradox or Russell's paradox, which all have this feature in common and of course, contain some special rules for the respective paradoxical connectives.
Ekman \citeyearpar{Ekman1994, Ekman1998} gives an example of a derivation of $\bot$, though, not containing any other rules than the usual ones for implication but which still, if we accept Ekman-reduction that is, could not be brought into normal form because as with Tennant's examples the reduction sequences are looping.
Thus, he concludes that Tennant's criterion does not capture a genuinely \textit{paradoxical} feature of the derivations considered, since with Ekman-reduction we could get such a derivation, as well, without containing any paradoxical elements.
There have been attempts to show that this can be avoided by using a different representation of the rules, e.g., in \citep{Plato2000} by using general elimination rules in ND showing that such a derivation can be brought to a normal form or in \citep{Tennant2021} with rules in sequent calculus showing that we get a cut-free derivation in such a system.
This does not stand in opposition to what I am focussing on in this paper, though.
Note that these `solutions' to the so-called \emph{Ekman-paradox} do concede that Ekman-reduction is permissible, since only by using it, we get into this infinite loop of reduction sequences.
If we reject Ekman-reduction for independent reasons, for which I will argue in this paper, then Ekman-paradox is no problem either.

The problem with this reduction, which Schroeder-Heister and Tranchini \citeyearpar{PSHT2017} point out and neatly prove, is the following: if we allow Ekman-reduction (plus always assuming for now that proofs related via reductions can be identified), then we would be forced to identify every derivation of a formula with every other derivation of the same formula, i.e., there would be no basis to distinguish different derivations other than their obvious syntactic difference.
We would have to commit to them all representing one and the same proof.
It is on these grounds of proof identity that Schroeder-Heister and Tranchini argue that Ekman-reduction should not be counted as an acceptable reduction.
I want to show now that annotating Ekman-reduction with terms shows something else that is essentially problematic with this reduction.

In our term-annotated system the derivation to which Ekman-reduction is applied is the following:\footnote{Note that Schroeder-Heister and Tranchini also consider a more general form of this reduction in their paper in that $A \rightarrow B$ and $B \rightarrow A$ are not assumptions but are derived formulas. Since this would not change the results here, I will stick to the original form, though.}

\quad \quad \quad  
\infer[\scriptstyle\rightarrow E]{App(y, App(x, t)): A} 
{\;\;\;\;\;\;\;\;\;{y: B \rightarrow A} \quad \infer[\scriptstyle\rightarrow E]{App(x, t) : B}{\;\;{x: A \rightarrow B} \quad \quad {\infer*{t:A}{\mathcal{D}}}}}

\vspace{0.3cm}

So the Ekman-reduction procedure for terms would be:

\begin{description}
\item[\textbf{Ekman-reduction:}] $App(y, App(x, t)) \rightsquigarrow_{Ekman} t$
\end{description} 
In the specific case above this reduction seems fine.
However, the problem with it, as opposed to the known $\beta$-reductions, is that it is too unspecific concerning the term structure.
With the same terms (since in Curry-style the types are not part of the syntactical structure of the terms) the following derivation could be constructed:\footnote{Still, there are reasons why Curry-style typing is preferable to Church-style, see section 5.}

\quad  
\infer[\scriptstyle\rightarrow E]{App(y, App(x, t)): A} 
{\;\;\;\;\;\;\;\;\;\;\;\;\;\;\;\;{y: B \rightarrow A} \quad \infer[\scriptstyle\rightarrow E]{App(x, t) : B}{\;\;{x: (A\rightarrow A) \rightarrow B} \quad \quad {\infer*{t:A \rightarrow A}{\mathcal{D}}}}}

\vspace{0.3cm}

The derivation is fine but the reduction would be problematic.
In this case it is clear that $A \not\equiv A \rightarrow A$, i.e., it cannot be the case that the term for $A$ and the term for $A \rightarrow A$ constructed out of the same type context in $\mathcal{\mathcal{D}}$ are syntactically the same.
Using a concrete example, we can show why allowing this reduction can create a problem.
Consider the following derivation:

\vspace{0.2cm}

\quad  
\infer[\scriptstyle\rightarrow E]{App(y, App(x, \lambda z.z)): \rho} 
{\;\;\;\;\;\;\;\;\;\;\;\;\;\;\;\;{y: \tau \rightarrow \rho} \quad \infer[\scriptstyle\rightarrow E]{App(x, \lambda z.z) : \tau}{\;\;{x: (\sigma \rightarrow \sigma) \rightarrow \tau} \quad \quad {\infer[\scriptstyle\rightarrow I]{\lambda z.z: \sigma \rightarrow \sigma}{[z:\sigma]}}}}
\vspace{0.3cm}

So, $App(y, App(x, \lambda z.z))$ would Ekman-reduce to $\lambda z.z$. 
However, no type assigned to $\lambda z.z$ can be atomic, i.e., $\lambda z.z : \rho$ is impossible.
The problems arising for these reductions relate to questions of so-called \emph{type preservation}, \emph{typechecking}, and \emph{type reconstruction}, which I will discuss in the next section.

\subsection{Subject reduction and type reconstruction}

The problem of \texttt{tonk}- and Ekman-reduction seems to be that, unlike the $\beta$-reductions, they are not type preserving.
Let us briefly take a look at this property and its significance for reduction procedures. 
Sometimes the expressions \textit{subject reduction} and \textit{type preservation} are used synonymously.
However, type preservation describes a broader concept than subject reduction, since the latter only says that types are preserved when terms (i.e., ``subjects") are \textit{reduced}, whereas type preservation can also be used to describe a property of subject \textit{expansions}.
So, we will distinguish this terminology here.\footnote{If in the following a reduction is stated (not) to be type preserving, this means that it enjoys (no) subject reduction, i.e., reductions are to be understood in the one-directed sense without looking at the other direction of expansions.}
The subject reduction theorem for the proof system with $\lambda$-terms we consider states the following \cite[p. 59]{SU}: 
\begin{description}
\item[\textbf{Subject Reduction Theorem:}] If $\Gamma \vdash t: A$ and $t \rightsquigarrow _{\beta} t'$, then $\Gamma \vdash t': A$.
\end{description}

Subject expansion, on the other hand, does not hold for this system in general, i.e., it is not the case that if  $t$ $\rightsquigarrow _{\beta}$ $t'$ and $t':A$, then $t:A$, meaning that the set of types assigned to a term is not invariant under conversion in general (see, e.g., \cite[p. 41]{Barendregt1992}; \cite[p. 170]{HS2008} for counterexamples).

The examples given in the previous section clearly show that subject reduction does not hold for \texttt{tonk}- and Ekman-reduction, i.e., it is not the case that whenever $t:A$ and $t \rightsquigarrow_{Ekman/\texttt{tonk}} t'$, then $t':A$.
We can also say that the contractum does not \textit{typecheck} at every type the redex typechecks at.
Typechecking is something that needs to be considered in Curry-style type systems \cite[see, e.g.,][p. 60]{SU} and is about deciding whether or not $\Gamma \vdash t:A$ holds, for a given context $\Gamma$, a term $t$ and a type $A$. We can express typechecking in the following form then:
\begin{description}
\item[\textbf{Typechecking:}]
$t$ \emph{typechecks} at $A$ iff $\Gamma \vdash t:A$ holds, for a given context $\Gamma$, a term $t$ and a type $A$.
\end{description}

As can be seen above, there are cases with Ekman-reduction (and for \texttt{tonk}  it is even more obvious) in which it is \emph{impossible} to assign $t'$ the type assigned to $t$.
If we understand types like labels telling us the combinations that can safely be made with a term, then we can understand subject reduction as saying that a term will not become `less safe' during a reduction, i.e., when performing a computation on a term, this term cannot turn from a well-typed into an ill-typed one \cite[p. 168]{HS2008}.
Subject reduction thus establishes the correctness of our system of type assignment \cite[p. 59]{SU}.
It seems, therefore, that maintaining subject reduction would certainly be a desirable feature for reduction procedures.

So, is subject reduction a good criterion to measure the acceptability of reduction procedures?
To deal with this question we need to determine whether non-type-preserving reductions necessarily lead to trivialization of the system.
In other words, are there systems which can contain reduction procedures that are not type preserving but yet do not trivialize the reducibility relation?
Although it looks like a promising criterion for reductions,   it actually seems to be the case that failure of subject reduction need not necessarily cause trivialization.
To wit, it does not seem impossible that there could be a type theory with a reduction that is not type preserving without relating terms of arbitrary types but, e.g., only of types which are equivalent (i.e., interderivable formulas).\footnote{An anonymous reviewer pressed the point here that it seemed unnecessary to look for a weaker criterion than subject reduction since with it we do get rid of the problematic cases. However, this would only amount to give a sufficient criterion, not a necessary one. Indeed, if a reduction enjoys the property of subject reduction, then it will be deemed acceptable and the criterion I will propose here will not stand against that. But that does not exclude the possibility that there might be reductions that are acceptable without having this property (see section 3.4).}

 The actual problem with the \texttt{tonk}-/Ekman-reductions, though, leading to trivialization of the system, can be identified when looking at type reconstruction for their redexes.
Type reconstruction is used to decide the typability of terms \cite[p. 60]{SU}:

\begin{description}
\item[\textbf{Type Reconstruction:}] Given term $t$, decide if there is a context $\Gamma$ and a type $A$, such that $\Gamma \vdash t :A$.
\end{description}

This can be achieved using a type reconstruction algorithm, which is simply based on the type assignment rules that are used.
Since these are just given by the annotated inference rules of our system, they are syntax-oriented.
This means that we should be able to figure out the \emph{principal types} of terms, i.e., figure out the derivation by reconstructing bottom-up the term using the type assignment rules. 
To give an example of a successful type reconstruction, let us consider the one for the redex $App(\lambda x.t, s)$ resulting from our $\rightarrow$-rules starting with assigning it an arbitrary type $B$.
We write `?' whenever this part of the type is syntactically undetermined in this step of the reconstruction.
Two occurrences of `?' in the same step mean that, although their structure is undetermined, they must be filled in by the same type symbol.
In the next step we are to use a `fresh' type symbol for `?'.

\vspace{0.2cm}
\textbf{Type reconstruction for} $App(\lambda x.t, s)$\textbf{:}

\begin{center}

\quad  
\infer[\scriptstyle\rightarrow E]{App(\lambda x.t, s):B} 
{{\lambda x.t:? \rightarrow B} \quad \quad \infer*{s:?}{\mathcal{D'}} }
\quad \quad \quad \quad
\infer[\scriptstyle\rightarrow E]{App(\lambda x.t, s):B} 
{{\infer[\scriptstyle\rightarrow I]{\lambda x.t:A \rightarrow B}{\infer*[\scriptstyle\mathcal{D}]{t:B}{[x:A]}}} \quad \infer*{s:A}{\mathcal{D'}} }
\end{center}

As we can see, the type reconstruction proceeds in such a way that \emph{we have to} assign contractum $t$ the same type as the redex.
The structure of the redex and the connected type assignment rules lead to an exactly determined type reconstruction, which cannot `go wrong' concerning the relation between types of redex and contractum.

\subsection{Criterion for acceptable reductions}

The problem with allowing a reduction such as the one for \texttt{tonk}, however, can be shown by a type reconstruction of the non-normal term $k'kt$, assuming $k'kt \rightsquigarrow\textsubscript{\texttt{tonk}}~ t$ as a reduction, as motivated above.
If we assign $k'kt$ an arbitrary type $B$, then the only information this gives us for $kt$ is that its type must be of the form ``? \texttt{tonk} $B$''.
Consequently, $t$ can be assigned an arbitrary type.
This means that the types of redex and contractum are arbitrarily independent of each other, which is exactly the core of the problem with a reduction for \texttt{tonk}.

\vspace{1.5cm}
\textbf{Type reconstruction for} $k'kt$\textbf{:}

\begin{center}

\quad \quad \quad 
\infer[\scriptstyle tonk E]{k'kt:B}{kt: ?~ \texttt{tonk}~ B}
\quad \quad \quad 
\infer[\scriptstyle tonk E]{k'kt:B}{\infer[\scriptstyle tonk I]{kt: A~ \texttt{tonk}~ B}{{t:A}}}
\end{center}

With type reconstruction it also becomes evident that the same problem as with \texttt{tonk} prevails with Ekman-reduction.
 We are doing a type reconstruction for the redex again.
If we assign $App(y, App(x, t))$ an arbitrary type $A$, then we can reconstruct bottom-up the following derivation in which a new type variable is used whenever it is independent from the ones already used (skipping the step-by-step illustration with `?'):

\vspace{0.2cm}
\textbf{Type reconstruction for} $App(y, App(x, t))$\textbf{:}

\vspace{0.2cm}

\quad \quad \quad \quad 
\infer[\scriptstyle\rightarrow E]{App(y, App(x, t)): A} 
{\;\;\;\;\;\;\;\;\;{y: B \rightarrow A} \quad \infer[\scriptstyle\rightarrow E]{App(x, t) : B}{\;\;{x: C \rightarrow B} \quad \quad {\infer*{t:C}{\mathcal{D}}}}}

\vspace{0.3cm}

Again, such a reduction allows reducing a term of one type to one of an \emph{arbitrary} other; one that is arbitrarily unrelated in the type reconstruction from the type of the term that is reduced.

This arbitrariness cannot arise with the standard $\beta$-reductions and, importantly, there are also other non-standard reductions which are well-behaved with respect to this feature, i.e., this is not simply to say that $\beta$-reductions are the only acceptable reductions.
For instance, if we compare \texttt{tonk}-reduction to the Liar-reduction given above, of course, they look very similar.
But in the Liar case type reconstruction quickly shows that this reduction is well-behaved, while the \texttt{tonk}-reduction is not.

So, what we are actually asking for is what I will call a `weak' subject reduction:

\begin{description}
\item[\textbf{Weak Subject Reduction:}] 
\item (i)  If $\Gamma \vdash t:A$ and  $t \rightsquigarrow t'$, then $\Gamma \vdash t':A$, or 
\item (ii) if $\Gamma \vdash t:A$,  $t \rightsquigarrow t'$ and $\Gamma \not\vdash t':A$, then it is not the case that $\Gamma \vdash t':B$ for arbitrary $B$.
$B$ is considered \emph{arbitrary} iff the rules of type assignment do not determine the type reconstruction of $t$ in a way that $B$ is related to $A$.

\end{description}

This is what I propose to demand as a criterion for a reduction to be acceptable: it should enjoy the property of weak subject reduction.
To reformulate it in other words, what we demand is that for the case that `full' subject reduction, i.e., clause (i), fails, $\Gamma \vdash t': B$ holds only for those $B$, which the rules of type assignment relate to $A$ in the type reconstruction of $t$.
This criterion ensures that whenever subject reduction holds, weak subject reduction holds as well, i.e., failure of weak subject reduction also implies failure of `full' subject reduction. That is important because it means that not meeting this desideratum only rules out the `bad' reductions.
The ones for which `full' subject reduction holds, which, as we said above, is deemed to be sufficient for `acceptable' reductions, cannot be ruled out by that criterion.
Also, note that failure of weak subject reduction does not necessarily mean that there is something wrong with the rules of type assignment in question.
In the case of Ekman-reduction there is nothing wrong with the rules, since the only rules used are the ones for $\rightarrow$ and those are fine for the $\beta$-reduction.
It rather shows that the reduction generated on grounds of these rules is misbehaved: it may work for specific types but it cannot be generalized in the same way `proper' reductions can.

Our way of \emph{checking} whether weak subject reduction holds or not is then via type reconstructions in the way described above: We conduct a type reconstruction for a term that would count as non-normal under this reduction, i.e., a redex, choosing `fresh' types whenever the type assignment rules allow this.
If the resulting types of redex and contractum occurring in this reconstruction are of arbitrarily different, unrelated types, then weak subject reduction fails and this means that this reduction should be rejected.
Thus, we do not only have a clear criterion of what distinguishes acceptable from unacceptable reductions but also a fairly simple way of testing this by the respective type reconstruction.
That can be considered an advantage when comparing it to Schroeder-Heister and Tranchini's way of showing how Ekman-reduction leads to unacceptable consequences, which they do by giving a very well-thought-out example of certain derivations leading to these consequences.
This is a very clever and sophisticated way, for sure, but one has to be able to come up with these examples in the first place.
Here, on the other hand, we have a systematic procedure of checking whether a reduction is acceptable or not.

\subsection{Type theory of core logic - another problematic case?}
In the following I want to give a concrete example of a reduction which is not type preserving, i.e., does not enjoy `full' subject reduction, but still does not necessarily have to be dismissed as a `bad' reduction.\footnote{Another example could be found in \citep{Wansing1993}, where a type theory for Nelson's logic with strong negation, \texttt{N4}, is given, which identifies terms of type $A$ with terms of type $\sim\sim A$. A concrete reduction is not formulated there but it is likely that it would have similar features as the one discussed in this section, since in such a system there would have to be rules of type assignment for $\sim$ which in some way relate $A$ and $\sim\sim A$.}
The reduction is presented by Ripley \citeyearpar{Ripley2020b} as part of an interesting typed term calculus for Tennant's Core Logic, i.e., an intuitionistic relevant logic.
The calculus, called \textit{Core Type Theory}, is interesting because it displays some very unusual features, while at the same time it is - at least in some respects - quite well-behaved.

According to Ripley the system maintains a similar correspondence to the impli-cation-negation-fragment of core logic as the one established by the Curry-Howard correspondence between the simply typed $\lambda$-calculus and intuitionistic logic.
In the proof system that he presents, next to formulas and connectives we have \frownie , which is related to negation but should not be considered as something like $\bot$.
\frownie ~is neither a formula nor a connective, i.e., it cannot be used to form any complex formulas, but rather, it is understood as a ``structural marker that interacts with the connective rules'' in a way specified by the proof system \citep[p. 112]{Ripley2020b}.
One of the things to note is that in core type theory in addition to the usual case where terms are of certain types (served by the formulas), here terms can also have \frownie ~instead of a type, in which case Ripley speaks of \textit{exceptional terms}.
Also, Ripley uses Church-style typing, i.e., the types and \frownie~ are part of the syntax of the terms.

Ignoring differences in notation, redex and contractum are defined in the same way as above, i.e., $App(\lambda x. t,s)$ as redex and the corresponding term $t[s/x]$ as contractum.\footnote{It may be noted here that in the definition of those, the differences between Curry- and Church-style typing are blurred somehow because the terms used in the definition are not typed, which would be the usual thing to be done in Church-style, see \cite[p. 26]{Hindley1997} and  \cite[p. 13]{TS}. S{\o}rensen and Urzyczyn \citeyearpar{SU}, who Ripley refers to in this context, leave out the types in their definition, as well, however, they say themselves that their `Church-style' is actually ``halfway between the Curry style and the `orthodox' Church style'' \cite[p. 66]{SU}.}
Relevant for our purpose is that the reduction procedure fails to be type preserving because it can happen that a typed term, on which we perform a reduction procedure, has an exceptional term as contractum.
Also, the system is not confluent, which means that normal forms are not unique, i.e., two syntactically distinct terms in normal form do not necessarily belong to two distinct equivalence classes generated by this reduction.
The system is indeed trivializing in the sense that \textit{if} we assume proof identity via the equivalence relation induced by its reduction procedure, then every term would have to be identified with every other term \cite[p. 128]{Ripley2020b}.
What must be stressed, however, is that there is only one non-type-preserving direction that is possible, namely from typed to exceptional terms.
We cannot go from terms of one type to terms of another type or from exceptional terms to typed terms.
This is one of the preservation properties this system still has.
Another is that the reduction can never lead to new free variables, i.e., the set of free variables in a redex is a (possibly proper) superset of the free variables in its contractum \cite[p. 116]{Ripley2020b}.
This is the same as in the simply typed $\lambda$-calculus.

Since this is no system of type assignment but a typed system \`{a} la Church, the issue of type reconstruction can actually not be raised \cite[p. 66]{SU}.
However, it seems rather unproblematic to convert Ripley's system into a system in which types and \frownie~ are assigned to terms according to the inference rules that are given.
The rules and reduction for $\rightarrow$ (he additionally considers rules for negation) would then look like this:

\vspace{0.3cm}

\quad  
\infer[\scriptstyle\rightarrow I]{\lambda x.t: A \rightarrow B} 
{\infer*{t : B}{{[x: A]} }}
\quad  
\infer[\scriptstyle\rightarrow I!]{\lambda x.t: A \rightarrow B} 
{\infer*{t : \frownie}{{[x: A]} }}
\quad  
\infer[\scriptstyle\rightarrow E]{r[App(s, t)/y]: C} 
{\infer*{s: A \rightarrow B}{{\Gamma}} \quad \infer*{t : A}{{\Delta} }\quad \infer*{r : C}{{[y:B]} }}

\vspace{0.2cm}
\quad
$App(\lambda x.t, s) \rightsquigarrow_{\beta} t[s/x]$
\vspace{0.2cm}

Since the inference rules are not as determined as our standard rules,\footnote{As can be seen, we have two $\rightarrow$-introduction rules.} it is clear that type reconstruction\footnote{If we want to be very precise, we would have to speak of ``\emph{hat} reconstruction'', ``rules of \emph{hat} assignment'', etc. since this is Ripley's terminology for including both types and $\frownie$. For simplicity, though, and because the criterion of weak subject reduction would still be met under such a reformulation, we will stick to the usual terminology.} cannot be conducted in such a way that it yields a determinate result as with the standard rules.
Importantly, however, neither does it result in complete arbitrariness of the kind we have seen with Ekman-reduction or \texttt{tonk}.
What can happen indeed, is that due to the two $\rightarrow$-I rules, we have \textit{two possible} paths in the type reconstruction, but that's it:\footnote{Note that although in core logic we have a generalized form of the elimination rule for $\rightarrow$, the instance of this rule here is the usual Modus Ponens since Ripley defines a redex being of the form $App(\lambda x.t, s)$.}

\begin{center}

\quad  
\infer[\scriptstyle\rightarrow E]{App(\lambda x.t, s):B} 
{{\lambda x.t:? \rightarrow B} \quad \quad \infer*{s:?}{\mathcal{D'}} }
\quad \quad \quad \quad
\infer[\scriptstyle\rightarrow E]{App(\lambda x.t, s):B} 
{{\infer[\scriptstyle\rightarrow I/\rightarrow I!]{\lambda x.t:A \rightarrow B}{\infer*[\scriptstyle\mathcal{D}]{\textcolor{red}{t:B/\frownie}}{[x:A]}}} \quad \infer*{s:A}{\mathcal{D'}} }
\end{center}

The two paths are marked by the step in red. 
Everything else will be exactly the same, though.
Since the reduction is $App(\lambda x.t, s) \rightsquigarrow_{\beta} t[s/x]$, it can happen that the redex reduces to a contractum which does not have the same type (neither does it have \emph{another} type, though, because $\frownie$ is no type at all). 
So, this means that the reduction in this system is not type preserving, i.e., subject reduction fails.
However, \emph{weak} subject reduction holds since it cannot reduce to an arbitrary type.
The contractum will be either of the same type as the redex or it will be assigned $\frownie$, which \emph{is} related to $B$ by the type assignment rules: thus, to this extent the type reconstruction is determined.

Therefore, we have a reduction in this system which is at least partially well-behaved.
On the one hand, confluence and subject reduction fail and if we would like the equivalence relation induced by reductions to give us proof identity, the reduction in this type theory would certainly not be suitable, since it trivializes identity of terms.
On the other hand, type reconstruction can be conducted in an ordered manner without the possibility of yielding arbitrary results.
Thus, the cases in which subject reduction fails are not completely arbitrary concerning the types, since it is not possible, as oppposed to Ekman- and \texttt{tonk}-reduction, that a well-typed term reduces to a term of an arbitrarily different type.  
An anonymous reviewer raised doubts about the acceptability of this system because the identity of terms would be trivialized by the reductions, demanding that disallowing this should rather be our minimal criterion for the acceptability of a system.
Note here that we must distinguish between the equivalence relation induced by the reductions and the reduction relation itself.
While the former is certainly too permissive to be interesting for a philosophical interpretation of the proof theory, the latter can still be recognized to be at least so well-behaved that it does not lead to an Ekman-\texttt{tonk}-ish kind of trivialization, which is the kind we are worried about for reasons to be discussed in the following section.

\section{Philosophical implications: Reduction procedures and meaning of proofs}
One of Prawitz's most important conjectures in this context is that, since the reductions induce an equivalence relation and two derivations should be considered to represent the same proof iff they are equivalent, proofs relating via these reductions are identical in nature.\footnote{What is left undecided in Prawitz's remarks \citeyearpar[p. 257]{Prawitz1971} is whether the $\beta$-reductions are the only conversions preserving identity of proofs or whether expansion operations (corresponding to $\eta$-expansions) and permutative conversions for $\vee$-elimination and $\exists$-elimination should be considered as well. He seems to lean towards accepting at least the expansions, while Martin-L\"of \citeyearpar[pp. 100f.]{Martin1975} discards both kinds of operations for identity preservation. Girard \citeyearpar[pp. 16, 73]{Girard1989}, on the other hand, includes $\eta$-expansions but is highly sceptical w.r.t. the permutative conversions when it comes to the question of identifying the `real objects' represented by the ND derivations. Since I am concerned only with \textit{reductions} here and not conversions in general, I will leave this issue as it is.}
This means in general that one and the same proof may be linguistically represented by different derivations and that in natural deduction a derivation in \textit{normal form} is the most direct form of representation of its denotation, i.e., the represented proof object.\footnote{This Fregean formulation can be found in \citep{Tranchini2016}, where this is meant to explicate Prawitz's and Dummett's conceptions on these matters.}

Failure of weak subject reduction means to have reductions that relate terms of arbitrarily different types, i.e., proofs of arbitrarily different formulas.
If we consider reductions to induce identity of proofs, a feature that ultimately results in having to identify proofs of arbitrarily different formulas would certainly be undesirable.
However, there is no necessity to subscribe to this identity theory of proofs.
There are other views on theories about identity of proofs on the market, of course, or it is also possible to argue like Tennant \citeyearpar[p. S599]{Tennant2021}, who seems to be a bit of an agnostic when it comes to this question. He indicates, though, in response to the proposal made in \citep{PSHT2017} to discard Ekman-reduction because it leads to a trivialization of proof identity, that we do not know enough about identity of proofs to use it as a criterion for other conceptions.\footnote{Examples for other approaches to proof identity would be Stra\ss burger's \citeyearpar[e.g.,][]{Strassburger2019} based on graphical proof practice, like proof nets for linear logic or what he calls a `combinatorial' approach for classical logic. Another one is Wansing's \citeyearpar{Wansing2021} approach, where a notion of identity between derivations is defined on the basis of taking both the notion of proofs as well as \emph{disproofs} as primitive. On this account proofs of certain formulas can be identified with disproofs of other, in specific ways related formulas and based on this a bilateralist notion of synonymy between formulas is defined subsequently.}
However, reductions can also be conceived of as calculations, evaluations, or interpretations of the given program, as discussed in section 2.
I will argue here that if we go for the latter conception of reductions, failure of weak subject reduction still remains a problem, even if it is not problematic for the identity of proofs anymore.
While this would be a problem for the \emph{denotation} of proofs, I want to show that the arbitrariness is a problematic feature also concerning the \emph{sense} of proofs.

Tranchini \citeyearpar{Tranchini2016} argues that only proofs which contain connectives for which reduction procedures are available can have sense.
He bases his argumentation on the Prawitzian tradition that derivations in normal form can be identified with the proof objects, i.e., their denotation, and the fact that the reductions are the instruments with which we can bring a derivation to its normal form.
If reductions for terms are considered to be decisive for the meaning of proofs, it seems that we should be clear about the question of the present paper: What are the conditions of acceptable reduction procedures?
In \citep{Ayhan2021b} the general assumption from Tranchini is retained that the connectives appearing in a derivation need to have acceptable reductions in order for the derivation to have sense at all and based on this an approach with $\lambda$-term-annotated proof systems is motivated to spell out what the sense of derivations consists in.
It is argued that in a term-annotated setting the denotation of derivations is represented by the end-term\footnote{The term decorating the formula that is proven.} of the derivation in normal form, since this term encodes the ultimate proof.
The \emph{sense} of a derivation, on the other hand, consists in the \emph{set of terms occurring within the derivation} because those terms encode the intermediate steps in the construction of the complex end-term encoding the conclusion \cite[p. 578]{Ayhan2021b}.
Thus, these terms reflect the operations used in the derivation, i.e., they reflect the way that is taken to get to the denotation.
Since they determine how the end-term is built up, they can be seen as encoding a procedure, which, finally, yields the end-term.
This seems in accordance with what, e.g., Dummett \citeyearpar[pp. 232, 323, 636]{Dummett1973} (a ``procedure'' to determine the denotation), Girard \citeyearpar[p. 2]{Girard1989} (``a sequence of \emph{instructions}'') or Horty \citeyearpar[pp. 66-69]{Horty2007} (``senses as procedures'') say about Fregean sense (Girard even in the context of relating this to the ``proofs as programs" conception).\footnote{Of course, there are other approaches on the Fregean sense like Evans' \citeyearpar{Evans1982}, on whose conception the interpretation given here could not be considered since lacking denotation would mean lacking sense as well. However, I do not find that interpretation very convincing, especially not in this context but also not in general (see also \citep{sep-propositions} on the problems of this conception).}

According to Frege, what is crucial, is that the signs (here: the syntax) uniquely determine the sense and the sense uniquely determines the denotation. 
What can happen, though, is of course the classic example of Hesperus and Phosphorus, where there is the same underlying denotation but different senses attached to different syntax (i.e., different words).
In the context of proofs this would be indicated by different terms used within the derivations, ending, however, on the same end-term or being reducible to the same end-term.
The following example illustrates this with a derivation in non-normal form reducing to the other, which is in normal form, since $App(\lambda y.\lambda x.x, \lambda y.y) \rightsquigarrow_{\beta} \lambda x.x$:
\vspace{0.2cm}

\quad 
{\small \infer[\scriptstyle\rightarrow E]{App(\lambda y.\lambda x.x, \lambda y.y): p \rightarrow p}{{\infer[\scriptstyle\rightarrow I]{\lambda y.\lambda x.x: (q \rightarrow q) \rightarrow (p \rightarrow p)}{\infer[\scriptstyle\rightarrow I]{\lambda x.x: p \rightarrow p}{[x:p]}}} \quad {\infer[\scriptstyle\rightarrow I]{\lambda y.y: q \rightarrow q}{[y:q]}}} }
\quad \quad 
 \infer[\scriptstyle\rightarrow I]{\lambda x.x: p \rightarrow p}{[x:p]}

\vspace{0.2cm}

The relation of these conceptions and the (un)acceptability of reduction procedures is the following now.
Whether or not we see the reductions as generating identity, or `merely' in this directed way as calculations, makes a difference concerning the denotation but \emph{not} concerning the sense.
We could use the theory described here but only equate terms over $\alpha$-conversion, for example.
The derivations above, one reducing to the other, would not be identified anymore in this case but the senses would remain unchanged.
They would not be identified because the denotation is referred to by the end-terms and if we do not assume identity over $\beta$-reductions, then these terms could not be identified, i.e., they would point to different proof objects.
The senses, though, consist in both cases (whether or not we assume $\beta$-equality for the end-terms) of the terms occurring within the derivations, i.e., they are different from each other in both cases but each for itself does not change by that assumption about the denotation of the proofs.
It would still hold that the sense determines the reference, in that there cannot be one sense leading to different denotations, and, importantly, that the syntax determines the sense.
This can only be claimed to hold, however, if the rules (the syntax) determine the type reconstruction for the redexes of reductions (the sense) to the extent that types of redex and contractum are not arbitrarily unrelated.
Otherwise, it cannot be said that the syntax determines the sense.
That needs to be the case, though, since meaning must be governed by rules.
It cannot be arbitrarily generated.
So, even if we do not accept the assumption that reductions generate equivalence relations over which proofs can be identified, it still makes sense to disallow reductions which render the derivations they are related to meaningless.

\section{Remarks on possible objections and Church- vs. Curry-style typing}
An objection that may be raised against the present approach in general is that it does not actually tell us anything interesting or important about reductions but rather, that it shows a weakness of the underlying assumption that using such term-annotated proof systems is a good way to go.
The argument could be delivered along the following lines:
The version of giving Ekman-reduction in a term-annotated form, as presented in this paper, is too generalized to capture what the reduction in original form meant to express.
In the original form the assumptions $A \rightarrow B$ and $B \rightarrow A$ are clearly essential for having such a reduction but these disappear completely in the reduction for terms considered here.
Thus, it is inadmissible to take this reduction, lacking essential features, as a correspondence for the original Ekman-reduction and dismiss it on this basis.

Further, it could be argued that, in order to capture the original Ekman-reduction appropriately, a restriction on the types should be implemented.
Such a restriction could be: $App(y, App(x,t)) \rightsquigarrow_{Ekman} t$ iff $t$ typechecks at every type  $App(y, App(x,t))$ typechecks at.\footnote{Note that it does not suffice here to demand that there is a type $A$ such that both $App(y, App(x,t))$ and $t$ typecheck at $A$. To see why this is not enough, consider the example at the end of section 3.1. There is a type such that $App(y, App(x, \lambda z.z))$ and $\lambda z.z$ both typecheck at, namely $\sigma \rightarrow \sigma$ (if we had $\sigma \rightarrow \sigma$ instead of $\rho$ in the example, this would clearly work out).}
What you get thereby is basically subject reduction \emph{by definition}.
It still might seem a bit more generalized than the original Ekman-reduction because we do not demand that the types are the same but the only way they could differ now is in the variables used for the atomic formulas, i.e., the principal types are always the same.
Of course, with such a restriction we would not have the undesirable arbitrariness in type reconstruction and thus, such a reduction would be well-behaved.

However, such a restriction must be rejected as a `saviour' for Ekman-reduction, since implementing it would entirely beg the question of what we wanted to devise here.
Firstly, in exactly the same way \texttt{tonk}-reduction could be restored: By demanding that $k'kt \rightsquigarrow_{\texttt{tonk}} t$ iff $t$ typechecks at every type $k'kt$ typechecks at.
I do not see, though, that this is what we are philosophically interested in when we want to investigate the nature of reductions.
It could not be said anymore, as Prawitz did, that the reductions make the inversion principle explicit if what we are doing is to restrict them by definition to cases in which the inversion principle is maintained.
What we are interested in, concerning the question ``What are acceptable reductions?'', is not to decide case by case whether it makes a difference to eliminate a certain detour or not, but to have some generalized form about which we can make such a judgment.  

What must be considered, secondly, when asking why such a restriction should be rejected in our approach, is that by using it we would abandon basics of Curry-style typing and de facto do Church-style typing instead.\footnote{Whereas in Curry-style the syntax of the terms is independent of types, in Church-style types are part of the syntax of terms.
This means that each variable is \emph{uniquely} typed and therefore, e.g., $\lambda x^{A}.x^{A}$ is a term of type $A \rightarrow A$ but not of, let's say, $B \rightarrow B$.
In Curry-style, on the other hand, $\lambda x.x$ is a term of type $A \rightarrow A$ \emph{for every} $A$.}
One could claim now, thereby raising another objection, that indeed, it simply would be better to use Church-style typing. 
The related objection would be to say that these features of type reconstruction just show that Curry-style typing has a severe disadvantage over Church-style, namely a looseness of the connection between terms and types, which makes it less beneficial for an approach like the present.
In other words, what all of this shows, is not that there is something wrong with certain reductions, but that our typing system is not helpful for this question and that we should rather use a type system \`{a} la Church (against which the first objection could not be raised anymore, either).

However, if we are interested in proofs from a philosophical, rather than merely technical, point of view, then Curry-style typing is preferable to Church-style.
In Church-style you \emph{will} get invariance of types under conversion but just because of the definition of the language, not because it is an interesting property.
All terms are typed, i.e., so are the $\beta$-redexes. 
Thus, $\beta$-reduction is restricted w.r.t. types and type changes are prevented \cite[p. 26]{Hindley1997}.
It is exactly these features of Church-style language then, which prevent us from asking philosophically interesting questions.
Because the types are part of the syntax of the terms and the typing rules are just part of the definition of the language, they cannot be used to answer questions about a more primitive, underlying language like ``Is this term typeable/meaningful?'', ``Can this term be assigned this or that type?'',... 
If you think of these questions as applied to proofs, these are philosophically interesting questions, though.
With Curry-style we have a language in which those can be asked, in Church-style they are prevented simply by the definition of the language.

\section{Conclusion}

What cannot be provided by our analysis here is an exhaustive list of properties that reductions need to have in order to count as `good' because this seems to depend on the role one wants a reduction to fulfill, which differs in the literature.
All we can do is to draw a distinct line of what \emph{unacceptable} reductions are:  reductions which do not enjoy the property of \textit{weak subject reduction}, that is, which yield the possibility of type reconstructions in which redex and contractum are arbitrarily independent of one another.
Further, it can be claimed that if `full' subject reduction fails, this does not necessarily need to lead to the exclusion of such a reduction.
It is a reason to be careful about identifying terms via the reduction, though.
Within the framework I outlined in this paper we have three kinds of reductions: firstly, the ones that are clearly well-behaved, like $\beta$-reductions. They have what seem to be very desirable features, like having the Church-Rosser-property, preserving types, etc.
Secondly, we have reductions which are clearly not well-behaved.
Those would be Ekman-reduction, or \texttt{tonk}-reduction, or any reduction which does not allow for a meaningful system because it arbitrarily connects terms of different types.
Thirdly, we have reductions in between those two categories.
These would be reductions like the example we saw in section 3.4; ones which may lack desirable features but are still well-behaved enough that they need not necessarily be excluded. 
Whether or not one wants to accept them, then depends on the underlying philosophical theory (e.g., about identity of proofs) one is subscribing to.

What I showed in this paper is that the question of what makes up acceptable reductions is neither trivial nor easy to answer in a positive way.
Thus, I make do with a negative answer, just like Schroeder-Heister and Tranchini do in their elaborate analysis of the topic in saying that acceptable reductions are not to yield an equivalence relation that trivializes the identity of proofs.
While I agree with their analysis, I aimed at going a step further and show that even if one does not agree with the underlying assumption that reductions induce an identity relation for proofs, there are certain reductions, like Ekman-reduction, which still have to be considered problematic.
The main point is that having to identify all proofs of the same formula is surely undesirable but it is all about the denotation.
However, if we have to commit to a notion of reductions according to which terms of a certain type reduce to terms of arbitrarily unrelated types, then such a system cannot be considered rule-generated anymore, and thus, not meaningful.


\medskip
\bibliographystyle{apacite}
\bibliography{ReferencesGeneral}

\end{document}